%% file: template.tex
\documentclass[a4paper]{article}
\usepackage{INTERSPEECH2022}
\usepackage{multicol,multirow}
\usepackage{url,cite}
\usepackage{hyperref}
\usepackage{xcolor}
\usepackage{tabularx,lipsum}
\hypersetup{
    colorlinks=true,
    linkcolor=purple,
    filecolor=magenta,      
    urlcolor=purple,
}

\newcolumntype{Y}{>{\centering\arraybackslash}X}

\title{SASV 2022: The First Spoofing-Aware Speaker Verification Challenge}
\name{Jee-weon Jung$^{*,1}$\thanks{$^{*}$These authors contributed equally to this work.}, Hemlata Tak$^{*,2}$, Hye-jin Shim$^3$, Hee-Soo Heo$^1$, Bong-Jin Lee$^1$,\\ Soo-Whan Chung$^1$, Ha-Jin Yu$^3$, Nicholas Evans$^2$, and Tomi Kinnunen$^4$}
\address{
  $^1$Naver Corporation, South Korea,\\
  $^2$EURECOM, Sophia Antipolis, France,\\
  $^3$School of Computer Science, University of Seoul, South Korea,\\
  $^4$University of Eastern Finland, Finland}
\email{sasv.challenge@gmail.com}

\begin{document}
\maketitle
\begin{abstract}
The first spoofing-aware speaker verification (SASV) challenge aims to integrate research efforts in speaker verification and anti-spoofing. 
We extend the speaker verification scenario by introducing {\em spoofed} trials to the usual set of target and impostor trials. 
In contrast to the established ASVspoof challenge where the focus is upon separate, independently optimised spoofing detection and speaker verification sub-systems, SASV targets the development of integrated and jointly optimised solutions.
Pre-trained spoofing detection and speaker verification models are provided as open source and are used in two baseline SASV solutions. 
Both models and baselines are freely available to participants and can be used to develop back-end fusion approaches or end-to-end solutions. 
Using the provided common evaluation protocol, 23 teams submitted SASV solutions. 
When assessed with target, bona fide non-target and spoofed non-target trials, the top-performing system reduces the equal error rate of a conventional speaker verification system from 23.83\% to 0.13\%.
SASV challenge results are a testament to the reliability of today's state-of-the-art approaches to spoofing detection and speaker verification. 
\end{abstract}

\section{Introduction}
Automatic speaker verification (ASV) systems determine whether or not an input utterance contains speech uttered by a given, known speaker. 
As one of the most efficient, convenient, natural and non-intrusive biometric characteristics, ASV has found widespread application, most notably, in telephony-based scenarios. 
Reliability is crucial and must be maintained not only in the face of target and impostor trials or utterances, but also so-called spoofed utterances, namely manipulated, synthesised or specially crafted utterances designed to deceive or manipulate the ASV system. 

The resistance of ASV systems to spoofing attacks has been studied only relatively recently and within the context of the ASVspoof initiative and associated challenge series~\cite{ASV2021challenge}.  
ASVspoof tackles the threat of spoofing attacks using countermeasures (CMs), separate detection sub-systems in the form of binary classifiers designed to distinguish between bona fide and spoofed inputs.
While there are other approaches~\cite{de2012evaluation,sizov2015joint}, 
CMs are usually combined with the ASV system in the form of a gate, the role of which is to detect and reject spoofed utterances such that they are never treated by the ASV sub-system. 

While the combination of CM and ASV sub-systems have the potential to improve security through increased robustness to spoofing attacks (rejection of spoofed trials), there is also the potential for degraded usability (rejection of bona fide target trials). 
An integrated approach to assessment is hence desirable and should gauge the impact of spoofing and CMs upon the ASV system.
Such a strategy implies that neither CMs nor ASV systems should be assessed independently from each other. Ultimately, there is but a single goal -- reliable ASV.

The minimum tandem detection cost function (min t-DCF)~\cite{kinnunen-tDCF-TASLP} metric has been developed with this vision and is one approach to the assessment of integrated CM and ASV sub-systems.
Given the potential dependence between CM and ASV sub-systems, and given the integrated approach to evaluation, it seems logical that the sub-systems themselves should also be jointly developed and optimised. This is not the case with ASVspoof, for which CMs are developed by challenge participants, while the ASV system is designed by the organisers. 
Herein lies the original goal of the new \textbf{S}poofing-\textbf{A}ware \textbf{S}peaker \textbf{V}erification (\textbf{SASV}) Challenge.\footnote{\url{https://sasv-challenge.github.io}}

SASV aims to promote the study of jointly optimised or fused CM+ASV solutions in addition to single, integrated solutions~\cite{jung2022sasv,zhang2022new,Zhang2022Norm,teng2022sa,kang22sasv,xuechen_2022_sasv,li2022backend}. 
The former is a more flexible and incremental approach to SASV. Fusion can be applied using current CM and ASV sub-systems and can hence exploit future advances in both.
With the addition of a new fusion model, however, complexity is increased, even if perhaps only modestly. %The potential of joint optimisation is also limited since a fusion model has access only to the representations or scores produced by the CM and ASV sub-systems. 
The fusion of two separate sub-systems is also somewhat at odds from the spirit of solutions to the \emph{single} task of reliable ASV.

Single, integrated solutions, envisaged on a longer-term horizon, represent more of step change and a more substantive SASV solution; they demand the design of entirely new solutions to the single problem of reliable ASV.
The development of single, integrated solutions is expected to be more challenging; it calls for the learning of a new latent space for the representation of both speaker identity and artefacts related to utterance authenticity (spoofing).
No matter what the solution strategy, the primary objective is progress in reliable ASV, where reliability implies resistance to both spoofed as well as bona fide non-target inputs.

In building upon long-established best practice in addition to the momentum generated through ASVspoof, the new SASV Challenge is built upon a common evaluation protocol and benchmarking framework, metrics, open source baseline solutions and use of publicly available ASVspoof data, specifically the 2019 Logical Access (LA) database~\cite{wang2020asvspoof}. 
The two different solution strategies are described in Section~\ref{sec:related}.
SASV datasets and new metrics are introduced in Sections~\ref{sec:database}.
Baseline sub-systems and solutions are described in Section~\ref{sec:baseline}.
SASV Challenge results and solution strategies are presented in Sections \ref{sec:results} and~\ref{sec:strategies}. 
Conclusions are presented in Section \ref{sec:conclusion}.

\section{Related Works}
\label{sec:related}
In this section, we describe prior, related work in two approaches to SASV: (i)~back-end fusion; (ii)~single, integrated models~\cite{alegre2013spoofing,khoury2014introducing, sahidullah2016integrated, todisco2018integrated,li2019multi,li2020joint,shim2020integrated,gomez2020joint}. 
Among the first category are decision-level, score-level, and embedding-level approaches to back-end fusion. 
Decision-level fusion includes various cascade approaches. 
Approaches to score-level fusion can be either parameter-free (e.g., score-sum ensemble) or parameter-driven (e.g., Gaussian mixture model) where both utilise separate scores from ASV and CM sub-systems~\cite{todisco2018integrated}.
Embedding-level fusion can also be achieved using a model operating upon embeddings that lie in different latent spaces~\cite{gomez2020joint,shim2020integrated,kanervisto2021optimizing}. 
Prior work includes Gomez-Alanis et al.~\cite{gomez2020joint} and Shim et al.~\cite{shim2020integrated} which both propose deep neural network (DNN)-based models to jointly optimise ASV and CM embeddings and hence produce single SASV scores and decisions.
Kanervisto et al.~\cite{kanervisto2021optimizing} reports a tandem solution to jointly optimise ASV and CM systems using reinforcement learning.
 
Single, integrated models have also been proposed.
These classify utterances into target and non-target classes where the latter comprises both traditional bona fide as well as spoofed non-target trials~\cite{sizov2015joint,li2019multi,li2020joint,zhao2022multi}. 
Sizov et al.~\cite{sizov2015joint} propose a two-stage PLDA approach to the joint optimisation of speaker and synthesis (spoofing) channel variations in an i-vector space. 
First, a PLDA model is trained using only embeddings extracted from bona fide speech. 
Then, a synthesis (spoofing) channel subspace is trained using only embeddings extracted from spoofed speech.
Zhao et al.~\cite{zhao2022multi} propose an integrated
spoofing-robust automatic speaker verification (SR-ASV) system. 
It uses a multi-task learning framework with max feature map activation~\cite{wu2018light} and residual convolutional blocks to extract discriminative embeddings and scores from task-specific, CM and ASV layers. 
None of the work described above was performed using common databases and protocols.

\section{Databases, protocols and metrics}
\label{sec:database}
Described in this section are the two publicly available databases used for the SASV challenge.
The VoxCeleb2~\cite{voxceleb2} database is used for the training of ASV sub-systems (see Section~\ref{ssec:sybsystems}). 
The ASVspoof 2019 LA database~\cite{wang2020asvspoof} is used for the training of CM sub-systems (see Section \ref{ssec:baseline}).

\subsection{VoxCeleb2}
The VoxCeleb2 database\footnote{\url{https://www.robots.ox.ac.uk/~vgg/data/voxceleb/vox2.html}} was collected by crawling online videos of celebrity interviews. 
The database is extracted from 150,480 unique videos with an average individual utterance length of 7.8 seconds. 
The development partition of the VoxCeleb2 database is used for the training of ASV sub-systems. 
It contains over 2,000 hours of data corresponding to 5,994 speakers (61\% male). 
The absence of spoofed utterances necessitates use of the VoxCeleb2 database in conjunction with additional databases containing spoofed utterances.

\subsection{ASVspoof 2019}
The ASVspoof 2019 LA database~\cite{wang2020asvspoof} is generated from the VCTK\footnote{\url{http://dx.doi.org/10.7488/ds/1994}} source database~\cite{vctk}, which includes speech data captured from 107 speakers (46 male, 61 female). 
It consists of disjoint train, development, and evaluation partitions. 
Each partition contains both bona fide and spoofed utterances.
The latter are generated using 19 state-of-the-art VC, TTS and hybrid TTS-VC attack algorithms (6 for the train and development partitions, 13 for the evaluation partition). 
Containing 25,380 utterances corresponding to 20 different speakers and both bona fide and spoofed utterances, the ASVspoof database can be used for the training of CM sub-systems but can also be used in conjunction with the VoxCeleb2 database for SASV research.

\subsection{SASV protocols}
While the SASV protocols\footnote{\url{https://github.com/sasv-challenge/SASVC2022_Baseline/tree/main/protocols}} exploit ASVspoof 2019 LA data, they are different to those used by participants of the ASVspoof challenge; they are not CM protocols and are, instead, ASV protocols or, more specifically, SASV protocols.
The latter involve three types of trial: (i)~target, bona fide trials uttered by the same speaker as enrolment utterance(s); (ii)~bona fide non-target trials uttered by a different speaker as enrolment utterance(s); (iii)~spoofed non-target trials containing speech which is either synthesised or converted in order to resemble the voice of the same speaker as enrolment utterance(s). 
Disjoint protocols are provided to challenge participants to support their development and evaluation of SASV solutions.

\subsection{Metrics}
\label{sec:metric}
SASV performance is assessed using the classical EER (SASV-EER) as the primary metric.
In keeping with~\cite{sahidullah2016integrated,todisco2018integrated,shim2020integrated}, the task hence remains one of binary classification: target vs non-target. 
We define {\em non-target} as the set of bona fide non-target (impostor) trials and spoofed non-target trials.
Without adequate countermeasures, both can cause increases in the false acceptance rate.
As depicted in Table~\ref{tab:eer_types}, two additional EER estimates serve as secondary metrics.  
The speaker verification EER (SV-EER) involves combinations of target trials and bona fide non-target trials whereas the spoofing EER (SPF-EER) involves combinations of target trials and spoofed non-target trials.
The SV-EER and SPF-EER are estimated using different subsets of the full protocol used in estimating the SASV-EER. 
As such, they reflect the reliability of the model under different, extreme conditions in which there are either no spoofed non-target trials or no bona fide non-target trials.

\begin{figure*}[t!]
  \centering
 \includegraphics[trim={0cm 0.3cm 0.2cm 0cm},clip,width=0.9\linewidth]{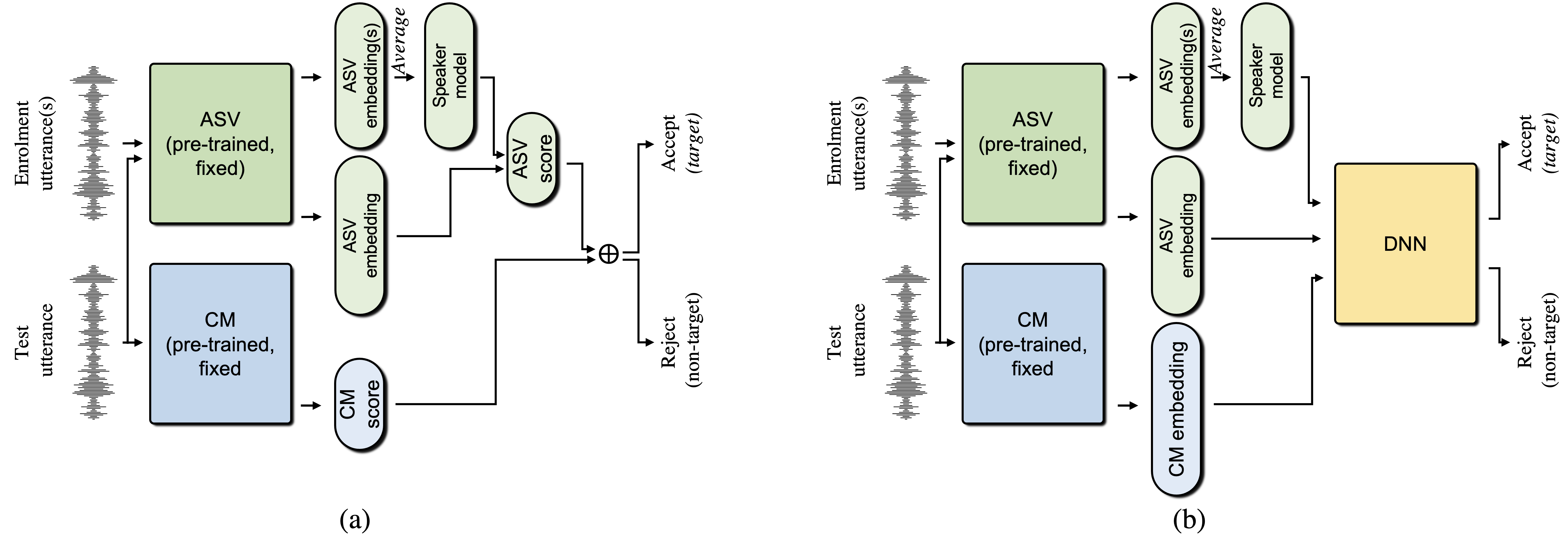}
 \caption{Illustration of the two baseline solutions of the SASV challenge. Both solutions utilise a pre-trained and fixed DNN-based ASV and CM sub-systems. (a): Baseline1 is a score-sum fusion where the cosine similarity score of ASV and CM output score are summed. (b): Baseline2 is a DNN-based fusion which involves speaker and spoofing embeddings.}
  \label{fig:Bsaeline2}
  \vspace{-10pt}
\end{figure*}

\input{tables/metrics}

\section{SASV baseline solutions}
\label{sec:baseline}

There are two baseline SASV systems. 
Both comprise standalone CM and ASV sub-systems.

\subsection{ASV and CM sub-systems}
\label{ssec:sybsystems}
%\newpara{ASV sub-system --}
ECAPA-TDNN is an efficient, robust state-of-the-art speaker embedding extractor~\cite{desplanques2020ecapa} 
consisting of a Res2net backbone architecture~\cite{gao2019res2net} with squeeze-excitation (SE) modules~\cite{hu2018squeeze}, a channel and context dependent statistics pooling layer and multi-layer feature aggregation to aggregate frame-level embeddings into utterance-level embeddings.
Inputs are 80-dimensional mel-filterbank features.
After aggregating frame-level representations, ASV embeddings are extracted using an affine transform
with a fully-connected (FC) layer. 
Data augmentation techniques are applied using the room impulse response database~\cite{rir} and additive noise recordings from the MUSAN database~\cite{musan}.
The network is trained using the recipe in~\cite{ecapatdnn_pretrained} and a reproducible open source implementation.\footnote{\label{fn:ecapa-tdnn}\url{https://github.com/TaoRuijie/ECAPATDNN}} Further details are available in~\cite{desplanques2020ecapa}.

AASIST~\cite{jung2022aasist} is an end-to-end state-of-the-art spoofing countermeasure system. 
It is based upon a RawNet2-based encoder~\cite{jung2020improved} and a spectro-temporal graph attention network (RawGAT-ST)~\cite{tak2021end}, novel heterogeneous graph attention layers and max graph operations to integrate temporal and spectral representations.
The output is generated using a readout operation and a hidden FC layer with two neurons.
CM embeddings of 160 dimensions are extracted prior to the FC output layer. 
AASIST is also available as open source.\footnote{\label{fn:assist}\url{https://github.com/clovaai/aasist}}
Further details are available in~\cite{jung2022aasist}.

\subsection{Baselines}
\label{ssec:baseline}
Open source baselines include: i)~\textbf{B1}: Score-sum fusion; and ii)~\textbf{B2}: DNN back-end fusion.\footnote{\label{fn:baseline}\url{https://github.com/sasv-challenge/SASVC2022_Baseline}} 
Both utilise pre-trained (ECAPA-TDNN\textsuperscript{\ref{fn:ecapa-tdnn}} and AASIST\textsuperscript{\ref{fn:assist}}) ASV and CM models. 
As illustrated in Figure~\ref{fig:Bsaeline2}-(a), baseline \textbf{B1} is a score-level back-end fusion method that combines ASV and CM sub-system outputs through score addition.
Since it requires neither additional training nor fine-tuning, back-end processing is parameter-free.
ASV scores are the cosine similarity between enrolment and test utterances.
A softmax non-linearity is optionally applied to the CM scores, which are otherwise unbounded, to normalise the scores within a (0,1) range.
The version of B1 with the softmax non-linearity is referred to as B1-v2.

As illustrated in Figure~\ref{fig:Bsaeline2}-(b),
baseline \textbf{B2} utilises an embedding-level fusion whereby a DNN operates upon a pair of speaker embeddings extracted from enrolment and test utterances, and a CM embedding extracted only from the test utterance.
The DNN is a vanilla multi-layer perception comprising three FC layers with leaky ReLU non-linear activation functions after each layer. 
The output layer consists of two neurons which correspond to the target and non-target (both bona fide and spoofed) classes. 
The model is trained using the training partition of the ASVspoof 2019 LA database. 
The larger-scale VoxCeleb2 database cannot be used for training since it contains only bona fide utterances.
Full details regarding challenge baselines are available in \cite{shim2022baseline}.

\input{tables/Challenge_results}

\section{Challenge results}
\label{sec:results}
From among 53 registrations, we received 23 submissions. 
SASV-EER results for each submission and the baselines for the evaluation partition are presented in Table~\ref{tab:challenge_results}.\footnote{\label{fn:result}
Full results are available at \url{https://sasv-challenge.github.io/challenge\_results}}
Without any countermeasure, the ECAPA-TDNN ASV sub-system gives an SASV-EER of 23.83\%.
Among the baseline solutions and with an SASV-EER of 1.71\%, baseline B1-v2 gives the best performance. 
With v2 having been developed after the release of the evaluation plan, in the remainder of this paper we emphasise v1 of the B1 baseline which gives an SASV-EER of 19.31\%, a result which serves to show the importance of score normalisation prior to fusion.
B2 gives an SASV-EERs of 6.54\%.

From among the 23 submissions, 21 outperformed the B1 and B2 baselines.
The submissions of 8 teams show SASV-EERs of below 1\%, a substantial reduction from that of the B1 baseline and is even below the SV-EER of 1.63\% (assessment without spoofed trials) of the ECAPA-TDNN ASV system~\cite{shim2022baseline}.
% In addition, SV-EER has higher correlation with the SASV-EER than the SPF-EER, although spoofed non-target trials outnumber bona fide non-target trials in the evaluation protocol.\textsuperscript{\ref{fn:result}}
These results are extremely encouraging and demonstrate the synergistic merit of combined ASV and CM systems.
The top-performing system achieves an SASV-EER of 0.13\% which corresponds to a 92\% relative reduction compared to the SV-EER of the ECAPA-TDNN ASV system. 

\begin{figure}[!t]
  \centering
  \includegraphics[width=0.9\linewidth]{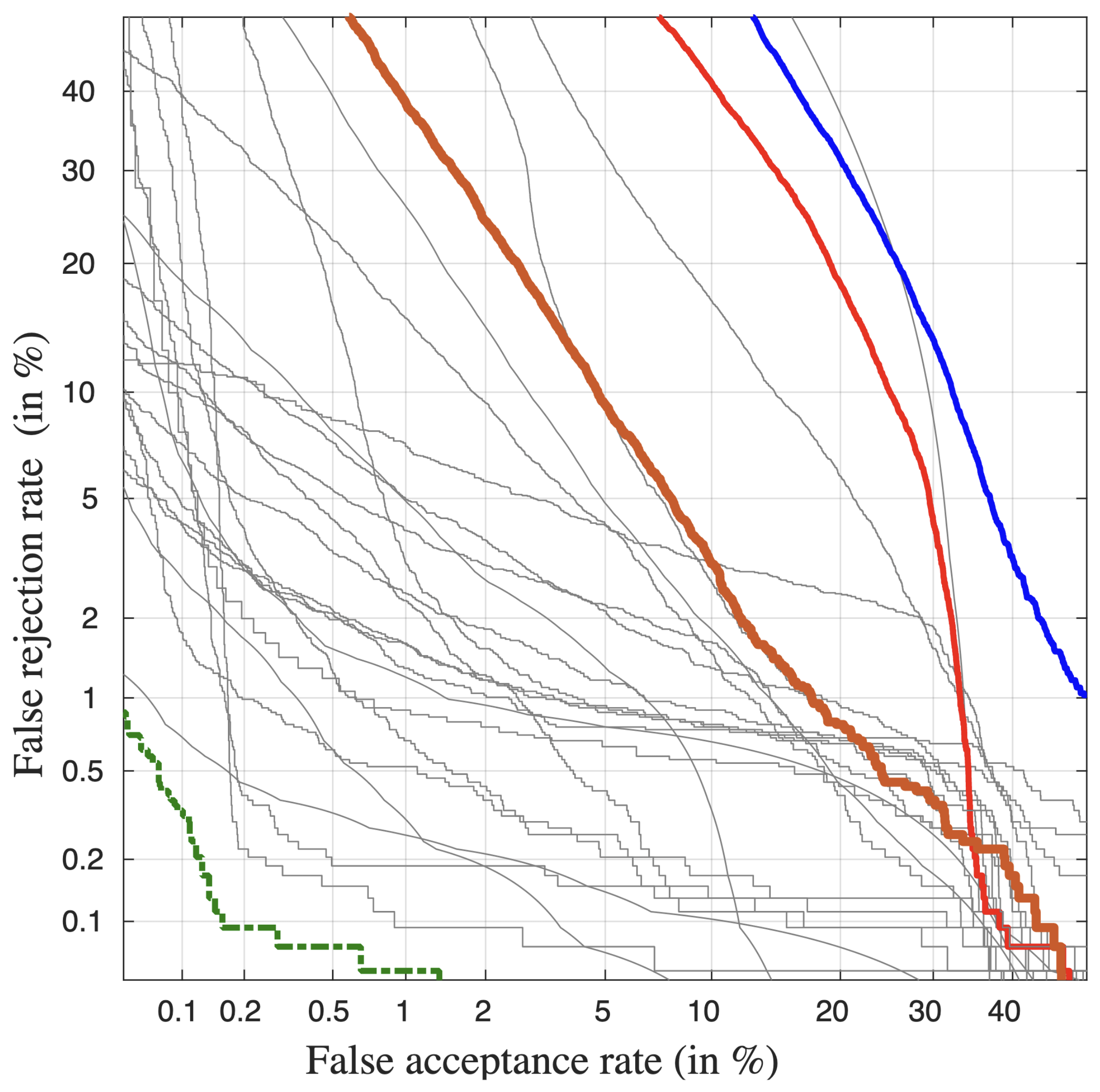}
  \caption{DET curves of challenge submissions on the evaluation protocol. {\em green}: best performing system,  {\em brown}: B2, {\em red}: B1 (without softmax), {\em blue}: ECAPA-TDNN ASV system.}
  \label{fig:DET_plots}
%   \vspace{-10pt}
\end{figure}

Detection error trade-off (DET) curves for all 23 submissions in addition to the baseline are plotted in Figure~\ref{fig:DET_plots}. 
Highlighted green, brown, red and blue profiles correspond to the top-performing submission and the B2, B1 baselines and ECAPA-TDNN ASV system respectively.
The plots show substantial differences in the trade-off between false rejections and false acceptances and point toward the potential for tuning combined ASV and CM solutions for operating points other than the SASV-EER.
Even so, the DET profile for the top-performing submission is below that of every other system.
The elbow that is visible in many profiles stems from assessment being performed using both bona fide and spoofed non-target trials.  
Even if SASV-EER results show higher correlation with SV-EER than with SPF-EER results,
%due to there being a greater number of spoofed non-target trials than bona fide non-target trails,
the SASV-EER is still a composite of the SV-EER and the SPF-EER.

\section{Solution strategies}
\label{sec:strategies}
All three systems with the lowest SASV-EER employed similar strategies~\cite{Alenin2022A,dku2022sasv}. 
First, they are all fusions of independent ASV and CM sub-systems.
Second, they are all ensembles of multiple ASV+CM systems. 
Third, each of the three systems gives similarly low SV-EERs and SPF-EERs indicating that they are insensitive to the proportion of spoofed non-target trials.
SV-EER and SPF-EER results for each system, for which we have no space to report here, are available online.\textsuperscript{\ref{fn:result}}

There are differences too. 
The IDVoice~\cite{Alenin2022A} system fuses ASV and CM sub-systems at the score-level. 
The DKU\_OPPO system performs fusion at the decision-level, whereas embedding level fusion is employed in the Hyu system.
Among their different systems, IDVoice use four different scores: ASV cosine similarity with and without score normalisation; CM cosine similarity; the score output of an end-to-end CM. 
Quality measurement functions~\cite{alenin2021id} are also used to normalise the score for each trial according to various meta information such as the duration of speech.
The DKU\_OPPO~\cite{dku2022sasv} system uses a modified cascade framework which combines decisions produced by the ASV sub-system with scores produced by the CM. 
Team Hyu focused on developing a latent space in which both speaker identity and spoofing artefacts can be captured using ASV and CM sub-system embeddings. 
A DNN with condition layers is used to map ASV and CM embeddings into a single SASV embedding~\cite{perez2018film}.
There are further differences in the training strategies which are described in
the participants' system descriptions and associated articles.\textsuperscript{\ref{fn:result}}

\section{Discussion and Conclusions}
\label{sec:conclusion}
The SASV Challenge was formed to promote the study of jointly optimised or fused and single solutions to reliable ASV.  
Reliability in this sense implies resistance to both bona fide and spoofed non-target trials.
We received submissions from 23 participating teams who all designed competing solutions using a common training and development protocol.
Results show substantial improvements over two Challenge baselines.
At 0.13\%, the lowest equal error rate for a spoofing-aware speaker verification system is even well below the equal error rate of a conventional, state-of-the-art speaker verification system assessed in the absence of spoofed trials.
This is an especially encouraging result which not only indicates the reliability of today's speaker verification and spoofing countermeasure technologies, but also the synergistic merit in their combination.
There is also evidence that such impressive levels of reliability are consistent no matter what fraction of non-target trials that are spoofed; similarly low SV-EER and SPF-EER results show that performance is insensitive to the spoofing prior.
We hope that these findings help to belay concerns of vulnerabilities to spoofing.

Looking to the future, we acknowledge some limitations of the current database.
All top-performing systems remain ensemble combinations of independent countermeasure and speaker verification sub-systems.
We anticipate future advances in joint optimisation stemming from the development of truly integrated, single system approaches.
Research in this direction will likely depend on the availability of larger databases containing not only spoofing attacks generated with diverse attack algorithms, but also data collected from \emph{many} more speakers.
We hope that these developments may improve yet further on what are already extremely encouraging results.

\section{Acknowledgements}
We thank members of the 23 teams who submitted results to this, inaugural edition.
We also thank advisory committee members, Héctor Delgado, Kong Aik Lee, Massimiliano Todisco, Md Sahidullah, and Xuechen Liu, for their help and support.

\clearpage
\bibliographystyle{IEEEtran}
\bibliography{mybib}
\end{document}

%% file: tables/metrics.tex
\begin{table}[t]
  \caption{Trial types used for estimation of each EER. ``+'' indicates the
positive class and ``-'' indicates the negative class. Each trial involves enrolment utterance(s) and a test utterance. 
The enrolment utterance(s) is bona fide (i.e. genuine) whereas test utterance belongs to either of the three types.}
  \centering
  \label{tab:eer_types}
  \begin{tabularx}{\linewidth}{lYYY}
    \toprule
    & \multirow{2}{*}{\textbf{Target}} & \textbf{Bona fide} & \textbf{Spoofed}\\
    &&\textbf{non-target}&\textbf{non-target}\\
    \toprule
    SV-EER & + & - &  \\
    SPF-EER & + &  & - \\
    SASV-EER & + & - & - \\
    \bottomrule
  \end{tabularx}
  \vspace{-15pt}
\end{table}

%% file: tables/Challenge_results.tex
\begin{table}[t!]
\caption{SASV 2022 Challenge results in evaluation SASV EER (\%). B1 and B2 correspond to the challenge baselines.} 
\vspace{-0.2cm}
%\renewcommand{\arraystretch}{1.1}
%team ID's illustrated in italics signify that use single integrated system for SASV task.}
\label{tab:challenge_results}
  \centering
  \setlength\tabcolsep{4.5pt}
  \centerline{
\begin{tabular}{clc|clc}
\toprule
\textbf{\#} & \textbf{Team ID} & \textbf{EER} & \textbf{\#} & \textbf{Team ID} & \textbf{EER}  \\ 
\toprule

1  & IDVoice     & 0.13 & 14 & VTCC & 1.86 \\ \hline
2  & DKU\_OPPO   & 0.21 & 15 & DeepASV & 2.48 \\ \hline
3  & Hyu         & 0.28 & 16 & SHELEZYAKA & 2.77 \\ \hline
4  & DoubleRoc   & 0.37 & 17 & xmuspeech & 2.89 \\ \hline
5  & FlySpeech   & 0.56 & 18 & HCCL & 4.30\\ \hline
5  & IRLAB       & 0.56 & 19 & magnum & 4.48\\ \hline
7  & VicomSpeech & 0.84 & 20 & CAU\_KU & 4.95\\ \hline
8  & CUHK\_NTU   & 0.89 & 21 &  Tandem & 6.22\\ \hline
9  & MARG        & 1.15 &    &B2& 6.54\\ \hline
10 & NII\_TJU    & 1.19 & 22 & DHU & 12.48 \\ \hline
11 & UR\_AIR     & 1.34 &    & B1&19.31\\ \hline
12 & clips       & 1.36 &    & ASV sub-system & 23.83\\\hline
13 & orange\_Lium& 1.48 & 23 & Souvik & 24.32\\\hline
 & B1-v2 & 1.71\\
\bottomrule

\end{tabular}}
\vspace{-13pt}
\end{table}

% 1 & IDVoice & 0.13 & 14 & VTCC & 1.86 \\ \hline
% 2 & DKU\_OPPO & 0.21 & 15 & DeepASV & 2.48 \\ \hline
% 3 & Hyu & 0.28 & 16 & SHELEZYAKA & 2.77 \\ \hline
% 4 & DoubleRoc & 0.37 & 17 & xmuspeech & 2.89\\ \hline
% 5 & FlySpeech & 0.56 & 18 & HCCL & 4.30\\ \hline
% 5 & IRLAB & 0.56 & 19 & magnum & 4.48\\ \hline
% 7 & VicomSpeech & 0.84 & 20 & CAU\_KU & 4.95\\ \hline
% 8 & CUHK\_NTU & 0.89 & 21 & Tandem & 6.22\\ \hline
% 9 & MARG & 1.15 & 22 & B2 & 6.54 \\ \hline
% 10 & NII\_TJU & 1.19 & 23 & DHU & 12.48\\ \hline
% 11 & UR\_AIR & 1.34 & 24  & B1 & 19.31\\ \hline
% 12 & clips & 1.36 & 25 & Souvik & 24.32\\ \hline
% 13 & orange\_Lium & 1.48\\